\documentclass[preprint,authoryear]{elsarticle}

\usepackage{color}
\usepackage{soul}
\usepackage{amsmath}
\usepackage{float}
\usepackage{booktabs}
\usepackage{setspace}
\usepackage{commath}
\usepackage[english]{babel}
\usepackage{multirow}
\usepackage[table,xcdraw]{xcolor}
\usepackage{subfig}
\usepackage{xfrac}
\usepackage{graphicx,epstopdf} 
\usepackage{lineno,hyperref}
\usepackage{moreverb} 
\usepackage{lipsum}

\makeatletter
\def\ps@pprintTitle{%
	\let\@oddhead\@empty
	\let\@evenhead\@empty
	\def\@oddfoot{}%
	\let\@evenfoot\@oddfoot}
\makeatother
\modulolinenumbers[1]

\begin{document}
%
%
%
%
%
%
%
%
%
%
%
%
%
%
%
%

\newpage 
\begin{frontmatter}
\title{Predicting kinetics using musculoskeletal modeling and inertial motion capture}
\author[xsens,utwente]{Angelos Karatsidis\corref{mycorrespondingauthor}}
\cortext[mycorrespondingauthor]{Corresponding author}
\ead{angelos.karatsidis@xsens.com}
\author[anybody]{Moonki Jung}
\author[xsens]{H. Martin Schepers}
\author[xsens]{Giovanni Bellusci}
\author[aau-hst]{Mark de Zee}
\author[utwente]{Peter H. Veltink}
\author[aau-mtech]{Michael Skipper Andersen}
\address[xsens]{Xsens Technologies B.V., Enschede 7521 PR, The Netherlands}
\address[utwente]{Biomedical Signals \& Systems, University of Twente, Enschede 7500 AE, The Netherlands}
\address[anybody]{AnyBody Technology A/S, Aalborg 9220, Denmark}
\address[aau-hst]{Department of Health Science and Technology, Aalborg University, Aalborg 9220, Denmark}
\address[aau-mtech]{Department of Materials and Production, Aalborg University, Aalborg 9220, Denmark \vspace{-1.0cm}}

\begin{abstract}
Inverse dynamic analysis using musculoskeletal modeling is a powerful tool, which is utilized in a range of applications to estimate forces in ligaments, muscles, and joints, non-invasively. To date, the conventional input used in this analysis is derived from optical motion capture (OMC) and force plate (FP) systems, which restrict the application of musculoskeletal models to gait laboratories.  To address this problem, we propose a musculoskeletal model, capable of estimating the internal forces based solely on inertial motion capture (IMC) input and a ground reaction force and moment (GRF\&M) prediction method. We validated the joint angle and kinetic estimates of the lower limbs against an equally constructed musculoskeletal model driven by OMC and FP system. The sagittal plane joint angles of ankle, knee, and hip presented excellent Pearson correlations ($\rho=0.95$, $0.99$, and $0.99$, respectively) and root-mean-squared-differences (RMSD) of $4.1\pm1.3^{\circ}$, $4.4\pm2.0^{\circ} $, and $5.7\pm2.1^{\circ}$, respectively. The GRF\&M predicted using IMC input were found to have excellent correlations for three components (vertical: $\rho=0.97$, RMSD=$9.3\pm3.0$ \%BW, anteroposterior: $\rho=0.91$, RMSD=$5.5\pm1.2$ \%BW, sagittal: $\rho=0.91$, RMSD=$1.6\pm0.6$ \%BW*BH), and strong correlations for mediolateral ($\rho=0.80$, RMSD=$2.1\pm0.6$ \%BW ) and transverse ($\rho=0.82$, RMSD=$0.2\pm0.1$ \%BW*BH). 
The proposed IMC-based method removes the complexity and space-restrictions of OMC and FP systems and could enable applications of musculoskeletal models in either monitoring patients during their daily lives or in wider clinical practice.
\end{abstract}

\end{frontmatter}

\begin{figure}[H]	
	\begin{minipage}[c]{0.2\textwidth}
		\href{http://creativecommons.org/licenses/by-nc-nd/4.0/}{\includegraphics[scale=0.7]{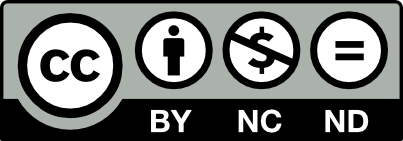}}		
	\end{minipage}\hfill
	\begin{minipage}[c]{0.78\textwidth}
		\caption*{\href{http://creativecommons.org/licenses/by-nc-nd/4.0/}{{\textcopyright} 2017. This manuscript version is made available under the CC-BY-NC-ND 4.0 license http://creativecommons.org/licenses/by-nc-nd/4.0/}
		} \label{cc-by-nc-nd}
	\end{minipage}
\end{figure}
\section{Introduction}
Assessment of muscle, joint, and ligament forces is important to understand the mechanical and physiological mechanisms of human movement. To date, the measurement of such in-vivo forces is a challenging task. For this reason, computer-based musculoskeletal models have been widely used to estimate the variables of interest non-invasively \citep{Damsgaard20061100,Delp20071940}.
\par 
The most common approach used in musculoskeletal modeling is the method of the inverse dynamics \citep{Erdemir2007}. This analysis utilizes the equations of motion with input from human body kinematics  in conjunction with kinetics obtained from external forces \citep{Winter1990}, to estimate joint reaction and muscle forces, as well as net joint moments using muscle recruitment methods \citep{Rasmussen2001409}. Measurements of the external forces are typically required and measured using force plates (FPs), however, the use of FPs has several limitations. First, subjects tend to alter their natural gait patterns in order to hit the small and fixed measurement area of a plate \citep{Challis200177}. In addition, this static and limited measurement area, impedes the assessment of several consecutive steps, when only a couple of FPs are available. Finally, the combined use of FP with motion input introduces a dynamic inconsistency, which results to residual forces and moments in the inverse dynamics. \citep{Riemer2008578,Hatze2002109}.
\par
Several studies have proposed replacing the FP input with wearable devices such as shoes with three-dimensional force and torque sensors beneath the sole \citep{Veltink2005423,Schepers2007895,Liu2010}. In a similar fashion, pressure insoles were proposed to reconstruct the complete ground reaction forces and moments (GRF\&M) from pressure distributions \citep{FornerCordero20041427,Rouhani2010311, Jung20142693}. Although these wearable devices are suitable for the assessment of external forces, the increased height and weight of the shoes equipped with force/torque sensors \citep{VanDenNoort20111381,Liedtke200739},  as well as the repeatability of the pressure sensors \citep{Low2010664} are considered important limitations.
\par
Recent research has suggested the replacement of the force input with predictions derived solely from motion input \citep{Audu20071115,Ren20082750,Oh20132372,Choi2013475,Fluit20142321,Skals2016}. In these studies, human body kinematics are combined with the inertial properties of the body segments, from which Newton-Euler equations are utilized to compute the external forces and moments. Since the system of equations becomes indeterminate during the double stance of gait, each of the aforementioned studies focused on methods to solve this issue.
\par
The majority of the existing research which studied the prediction of GRF\&M, used conventional optical motion capture (OMC) input. Despite the high accuracy of this method in tracking marker trajectories, its dependence on laboratory equipment restricts possible applications during daily life activities or in wider clinical practice. In the previous decade, ambulatory motion tracking systems based on inertial measurement units (IMUs), have been proposed as a suitable alternative for estimating 3D segment kinematics \citep{Luinge2005273,Roetenberg2005395,Roetenberg2009, Zhang2013}. A key benefit of such systems is that they can be applied in virtually any environment without depending on external infrastructure, such as cameras. Driven by these advances in inertial motion capture (IMC), recent work of the authors demonstrated its ability to estimate three-dimensional GRF\&M \citep{Karatsidis2017}, which were distributed between the feet using a smooth transition assumption concept \citep{Ren20082750}.
\par
To date, the use of musculoskeletal modeling with kinematic inputs from IMUs to estimate internal joint moments, muscle forces and joint reactions has only received limited attention. \citet{Koning20151003} previously demonstrated the feasibility of kinematically driving a musculoskeletal model using only orientations from IMUs. However, that study only compared the kinematics of the musculoskeletal model, without any inverse dynamic calculations. 
\par
The aim of this study was to drive a musculoskeletal model and perform full-body inverse dynamics using exclusively IMC data. We validate the kinematics as well as the predicted GRF\&M, joint reaction forces and net moments (JRF\&M) compared against an OMC and FP-driven musculoskeletal model. 
\section{Methods}
\subsection{Subjects}
The experimental data was collected at the Human Performance Laboratory, at the Department of Health Science and Technology, Aalborg University, Aalborg, Denmark following the ethical guidelines of The North Denmark Region Committee on Health Research Ethics. Eleven healthy male individuals with no present musculoskeletal or neuromuscular disorders volunteered for the study (age: 31.0 $\pm$ 7.2 years; height: 1.81 $\pm$ 0.06 m; weight: 77.3 $\pm$ 9.2 kg; body mass index (BMI): 23.6 $\pm$ 2.4 $\text{kg/m}^\text{2}$).  All participants provided written informed consent, prior to data collection. 
\subsection{Instrumentation}
Full-body IMC data were collected using the Xsens MVN Link (Xsens Technologies B.V., Enschede, the Netherlands), in which 17 IMUs were mounted on the head, sternum, pelvis, upper legs, lower legs, feet, shoulders, upper arms, forearms and hands using the dedicated clothing. The affiliated software Xsens MVN Studio 4.2.4 was used to track the IMU orientations with respect to an earth-based coordinate frame\citep{Luinge2005273,Roetenberg2005395}. Segment orientations were obtained by applying the IMU-to-segment alignment, found using a known upright pose (N-pose) performed by the subject at a known moment in time, while taking specific care to minimize the effect of magnetic disturbances. In addition, this information is fused with updates regarding the joints and external contacts to limit the position drift \citep{Roetenberg2009}.  
\par
For validation purposes, an OMC system utilizing 8 infrared high speed cameras (Oqus 300 series, Qualisys AB, Gothenburg, Sweden) and the software Qualisys Track Manager 2.12 (QTM) were used to track the trajectories of 53 reflective markers mounted on the human body, as described in the Appendix of \citep{Karatsidis2017}. In addition, three FP systems (AMTI OR6-7-1000, Advanced Mechanical Technology, Inc.,Watertown, MA, USA) embedded in the floor of the laboratory, were utilized using QTM to record the GRF\&Ms. Both IMC and OMC systems sampled data at a frequency of 240 Hz, while the FP system sampled data at 2400 Hz. A second-order forward-backward low-pass Butterworth filter was applied to the reflective marker trajectories and measured GRF\&M, with cut-off frequencies of 6 Hz and 15 Hz, respectively.

\subsection{Experimental protocol}
For each participant, the body dimensions were extracted using a conventional tape following the guidelines of Xsens. 
During the data collection, the subjects were instructed to walk barefoot in three different walking speeds (comfortable; CW, fast; FW, and slow; SW). The walking speeds performed experimentally were quantified as 1.28$\pm$0.14 m/s (mean $\pm$ standard deviation) for CW, 1.58$\pm$0.09 m/s for FW (CW $+$ 23\%) and 0.86$\pm$0.11 m/s for SW (CW$-$33\%). For every walking speed, five successful trials were assessed.  A successful trial was obtained when a single foot hit one of the FPs entirely, followed by an entire hit of the other foot on the successive FP.
\subsection{Overall description of the components in the musculoskeletal models}
Three musculoskeletal models have been constructed in AnyBody\textsuperscript{TM} Modeling System (AMS) v.6.0.7  (AnyBody\textsuperscript{TM} Technology A/S, Aalborg, Denmark) \citep{Damsgaard20061100}:
\begin{itemize}
  \item a model in which the kinematics are driven by IMC and the GRF\&M are predicted from the kinematics (IMC-PGRF).
  \item a model in which the kinematics are driven by OMC and the GRF\&M are predicted from the kinematics (OMC-PGRF).
  \item a model in which the kinematics are driven by OMC and the GRF\&M are measured from FPs (OMC-MGRF).
\end{itemize}

In the IMC-PGRF model, a Biovision Hierarchy (BVH) file is exported from Xsens MVN Studio and imported in AMS, in which a stick figure model is initially reconstructed. To match the stick figure model with the musculoskeletal model, we utilize a concept of virtual markers (VMs) \citep{Skals2017}. The VMs are mapped to particular points on each model, as illustrated in Figure \ref{fig:models} and described in the supplementary material. Following this step, the VMs are treated as actual experimental markers, as if they were derived from an OMC system. Contrary to OMC, no filtering was applied to the VM trajectories. 
\par
In all models, the GaitFullBody template of the AnyBody\textsuperscript{TM} Managed Model Repository (AMMR) 1.6.2 was used to reconstruct the musculoskeletal models in AMS. 
The lumbar spine model was derived from the study of \citet{deZee20071219}, the lower limb model was derived from the Twente Lower Extremity Model \citep{KleinHorsman2007239}, and the shoulder and upper limb models were based on the model of the Delft Shoulder Group \citep{Veeger1991615,Veeger1997647,VanderHelm1992129}. The full-body kinematic model contained 39 degrees-of-freedom (DOF) in total. Specifically, a pelvis segment with three rotational and three translational DOF, two spherical hip joints, two revolute knee joints, two universal ankle joints, a spherical pelvic-lumbar joint, two glenohumeral joints with five DOF each, two universal elbow joints, and two universal wrist joints. The motion of the neck joint was locked to a neutral position.  
\begin{figure}[ht]
\caption{Illustration of the pipeline used in the IMC-PGRF approach. A recording from Xsens MVN Studio (a) is exported to a BVH file to generate a stick figure model (b), in which virtual markers (blue) are placed. Virtual markers (red) are also placed on points of the musculoskeletal model (c), and by projecting b on c the kinematics of the musculoskeletal model are solved. Finally, inverse dynamic analysis using prediction of ground reaction forces and moments is performed to estimate the kinetics.}
\label{fig:models}
\includegraphics[width=\textwidth,height=\textheight,keepaspectratio]{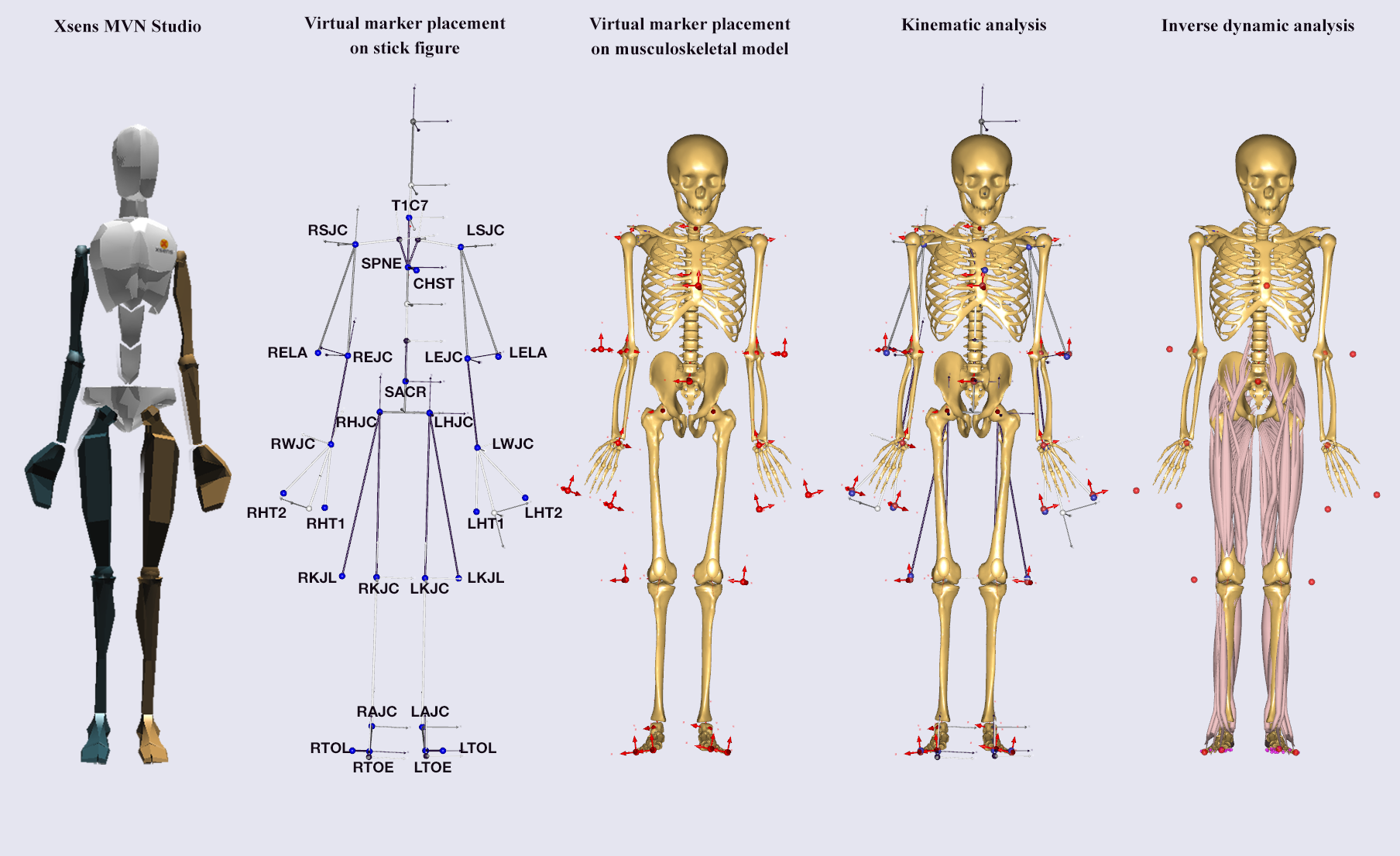}
\end{figure} 
\subsection{Scaling and kinematics analysis of the musculoskeletal models}
For each subject, a standing reference trial with an anatomical pose was utilized to identify the parameters of segment lengths and the (virtual) marker positions, using a least-square minimization between the model and input (virtual or skin-mounted) marker positions \citep{Andersen2010171}. In the IMC-PGRF musculoskeletal model, the lengths of the shanks, thighs, head, upper arm and forearms were derived directly from the stick figure, as generated from Xsens MVN studio using the measured body dimensions. In contrast, the pelvis width, foot length, and trunk height were optimized based on the above-mentioned least-square minimization method. The estimated segment lengths were used in all subsequent dynamic trials to perform the kinematic analysis based on the method of \citet{Andersen2009371}.  
\subsection{Inertial and geometric scaling of the musculoskeletal models}
The mass of each segment was linearly scaled based on the total body mass and the segment mass ratio values reported by \citet{Winter1990}. The inertial parameters were calculated by considering the segments as cylinders with uniform density. In addition, geometric scaling of each segment, where the longitudinal axis was defined as the second entry, was achieved using the following matrix:
\begin{equation}
    S = \begin{bmatrix}
        \sqrt{\frac{m_s}{l_s}} & 0 & 0 \\
        0 & l_s & 0 \\
        0 & 0 & \sqrt{\frac{m_s}{l_s}} 
     \end{bmatrix}    
\end{equation}
where $S$ is the scaling matrix, $l_s$ is the ratio between the unscaled and scaled lengths of the segment, $m_s$ is the mass ratio of the segment. 
\subsection{Muscle recruitment}
The muscle recruitment problem was solved by defining an optimization problem:
\begin{subequations}
\begin{equation}
\label{eq:costFunction}
\ G(\boldsymbol{f}^{(M)}) = \sum_{i=1}^{n^{(M)}}A_i\left(\frac{f_{i}^{(M)}}{N_i}\right)^3
\end{equation}
\begin{equation}
\label{eq:dynamicEquations}
\boldsymbol{Cf} = \boldsymbol{d}
\end{equation}
\begin{equation}
\label{eq:constraints}
0 \leq f_{i}^{(M)} \leq N_{i}, i =1,...,n^{(M)}
\end{equation}
\end{subequations}
This system of equations minimizes the cost function $G$ (\ref{eq:costFunction}), subject to the dynamic equilibrium equations (\ref{eq:dynamicEquations}) and non-negativity constraints, so that each muscle can only pull, but not push, while its force ($f_{i}^{(M)})$ remains below its strength ($N_i$) (\ref{eq:constraints}). The vector $\boldsymbol{f}$ contains all the unknown muscle and joint reaction forces, $\boldsymbol{f}^{(M)}$ denotes the muscle forces, and $n^{(M)}$ denotes the number of muscles. The physiological cross-sectional area of the $i$th muscle is denoted by $A_i$. The coefficient matrix $\boldsymbol{C}$ contains the equations of dynamic equilibrium and $\boldsymbol{d}$ the external loads and inertia forces \citep{Damsgaard20061100,Skals2017, Marra2015}. 
\par
The strengths of the muscles were derived from previous studies which described the models of the body parts, and were considered constant for different lengths and contraction velocities \citep{deZee20071219,KleinHorsman2007239,Veeger1991615,Veeger1997647,VanderHelm1992129}.
To scale the muscle strengths, fat percentage was used as in \citet{Veeger1997647}, calculated from the body mass index \citep{Frankenfield200126}. The model of the lower body contained 110 muscles, distributed into 318 individual muscle paths. In contrast, in the upper body model, ideal joint torque generators were utilized. Actuators for residual forces and moments with capacity up to 10 N and Nm, respectively, were placed at the origin of the pelvis and included in the muscle recruitment problem previously described.  
\subsection{Ground reaction force and moment prediction}
The GRF\&M were predicted by adjusting a method of \citet{Skals2016}. A set of eighteen dynamic contact points were overlaid 1 mm beneath the inferior surface of each foot. Each dynamic contact point consisted of five unilateral force actuators, which could generate a positive vertical force perpendicular to the ground, and static friction forces in the anterior, posterior, medial, and lateral directions using a friction coefficient of 0.5. In addition, the height and velocity activation thresholds were set to 0.03 m and 1.2 m/s, respectively.

\subsection{Data Analysis}
Lower limb joint angles calculated in the IMC-PGRF model were compared to the OMC-PGRF/OMC-MGRF. In addition, GRF\&M and JRF\&M of the IMC-PGRF and OMC-PGRF were compared to OMC-MGRF. 
\par
Forces were normalized to body weight (BW) and moments to body weight times body height (BW*BH). 
The time axis of the curves was normalized to 100\% of the gait cycle for the kinematics (time between two consecutive heel-strike events of the analyzed limb) and 100\% of the stance phase (time between heel-strike and toe-off events of the analyzed limb) for the kinetics. 
\par
The above-mentioned comparisons of kinematic and kinetic variables to their respective references were performed using absolute and relative root-mean-square-differences (RMSD and rRMSD, respectively)as described by \citet{Ren20082750}. In addition, for every curve, the magnitude ($M$) and phase ($P$) difference metrics \citep{Sprague2003149} have been utilized. Pearson correlation coefficient ($\rho$) were calculated, averaged using Fisher's z transformation method \citep{Silver1987averaging}, and categorized similarly to Taylor \textit{et al.} \citep{Taylor199035}, as "weak" ($\rho \leq 0.35$), "moderate" ($0.35 < \rho \leq 0.67$), "strong" ($0.67 < \rho \leq 0.90$), and "excellent" ($\rho > 0.90$). 

\section{Results}
\subsection{Estimated kinematics of the musculoskeletal model}
\begin{figure}[H]
\caption{Ankle, knee, and hip joint angle estimates (standard deviation around mean) of the IMC-PGRF (orange shaded area around orange dotted line) and OMC-PGRF models (blue shaded area around blue dashed line) versus OMC-MGRF model (thin black solid lines around thick black solid line).}
\label{fig:jointAngle}
\includegraphics[width=\textwidth,height=\textheight,keepaspectratio]{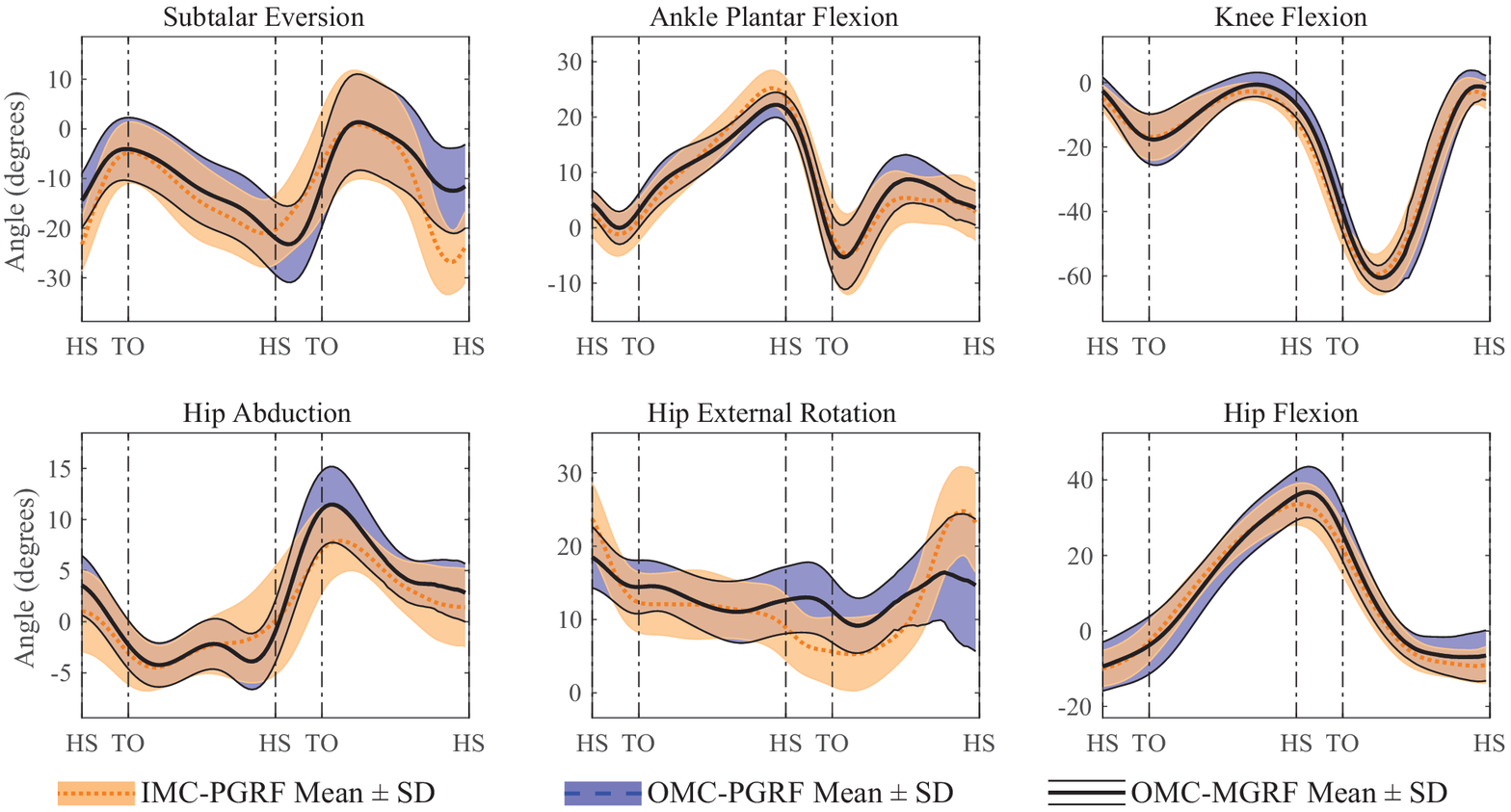}
\end{figure}

Table \ref{tab:IMCvsOMC} presents the results for the accuracy analysis for the joint angles of the IMC-driven model versus the OMC-driven model. Similarly, Figure \ref{fig:jointAngle} illustrates the curves for the joint angles of the lower extremities averaged across all gait cycles performed by the eleven subjects.
Excellent Pearson correlation coefficients have been found in all sagittal plane angles for ankle, knee, and hip ($0.95, 0.99$, and $0.99$, respectively). For the same variables, the RMSDs across a gait cycle were found as $4.1 \pm 1.3 ^{\circ}$, $4.4\pm 2.0 ^{\circ}$ and $5.7 \pm 2.1 ^{\circ}$, respectively (mean $\pm$ standard deviation). Hip flexion angles were overall underestimated ($M = -4.0 \pm 13.9\%$), whereas knee and ankle magnitude differences showed an average overestimation ($0.7\pm 6.2\%$ and $8.6 \pm 16.4\%$). 
The hip abduction showed excellent correlations ($\rho = 0.91$) with an RMSD of $4.1 \pm 2.0^{\circ}$ and a mean underestimation with a magnitude difference $M = -12.2 \pm 34.7\%$. Strong correlation values ($\rho=0.68$) were observed in the hip internal-external rotation angle with an RMSD of $6.5 \pm 2.8^{\circ}$ and an underestimation of magnitude difference $M=5.5 \pm 39.0\%$. Finally, the subtalar eversion angle showed strong correlation ($\rho = 0.82$), RMSD of $9.66 \pm 3.07^{\circ}$ and $M = 24.0 \pm 34.7\%$.

\begin{table}[t]
\centering
\footnotesize
\caption{Comparison of lower limb joint angles between musculoskeletal model driven by the inertial (IMC-PGRF) and optical motion capture (OMC-PGRF/OMC-MGRF), using Pearson correlation coefficient ($\rho$), absolute and relative root-mean-squared-differences ($RMSD$ in degrees and $rRMSD$ in \%, respectively). $M$ and $P$ denote the \% magnitude and phase differences .}
\label{tab:IMCvsOMC}
\begin{tabular}{lrrrrr}
\toprule
& \multicolumn{1}{c}{$\rho$} & \multicolumn{1}{c}{RMSD} & \multicolumn{1}{c}{rRMSD} & \multicolumn{1}{c}{M} & \multicolumn{1}{c}{P}  \\\midrule
Subtalar Eversion       & 0.81 & 9.7 (3.2) & 32.6 (10.3) & 24.0 (34.7)  & 19.3 (10.2) \\
Ankle Plantarflexion    & 0.95 & 4.1 (1.3) & 14.0 ( 4.8) & 8.6 (16.4)   & 9.8 ( 3.9)  \\
Knee Flexion            & 0.99 & 4.4 (2.0) & 7.2 (3.4)   & 0.7 (6.2)    & 4.8 (2.7)   \\
Hip Abduction           & 0.91 & 4.1 (2.0) & 25.9 (10.7) & -12.2 (34.7) & 21.2 (9.3)  \\
Hip External Rotation   & 0.68 & 6.5 (2.8) & 36.9 (15.2) & 5.5 (39.0)   & 12.6 (6.2)  \\
Hip Flexion             & 0.99 & 5.7 (2.1) & 12.7 ( 5.3) & -4.0 (13.9)  & 8.8 (4.2)  \\
\bottomrule
\end{tabular}

\end{table}

\subsection{Predicted kinetics using inertial and optical motion capture}
\begin{figure}[H]
\includegraphics[width=\textwidth,height=\textheight,keepaspectratio]{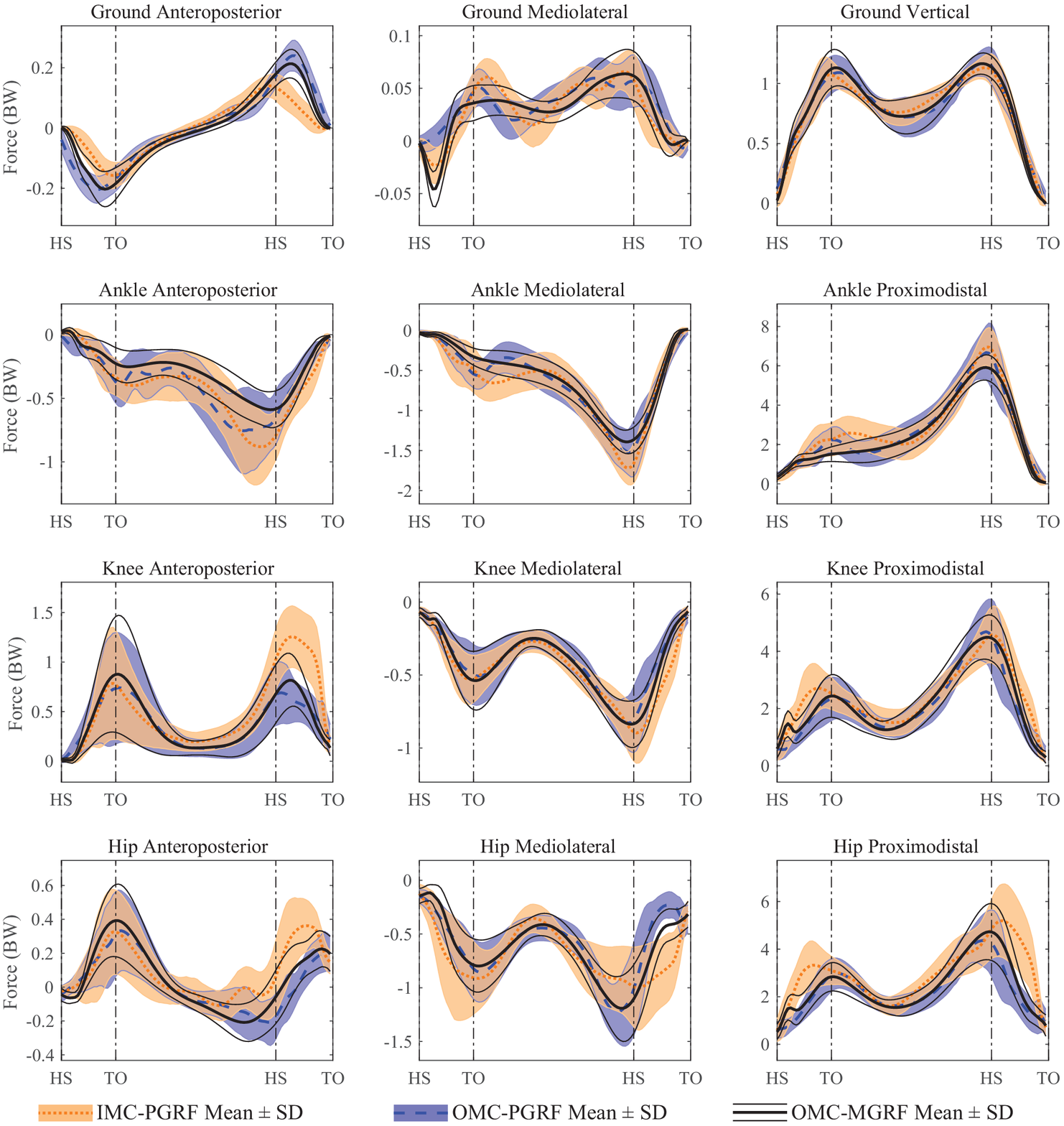}
\caption{Ground and lower limb joint reaction force estimates (standard deviation around mean) of the IMC-PGRF (orange shaded area around orange dotted line) and OMC-PGRF models (blue shaded area around blue dashed line) versus OMC-MGRF model (thin black solid lines around thick black solid line).}
\label{fig:Force}
\end{figure}

\begin{figure}[H]
\includegraphics[width=\textwidth,height=\textheight,keepaspectratio]{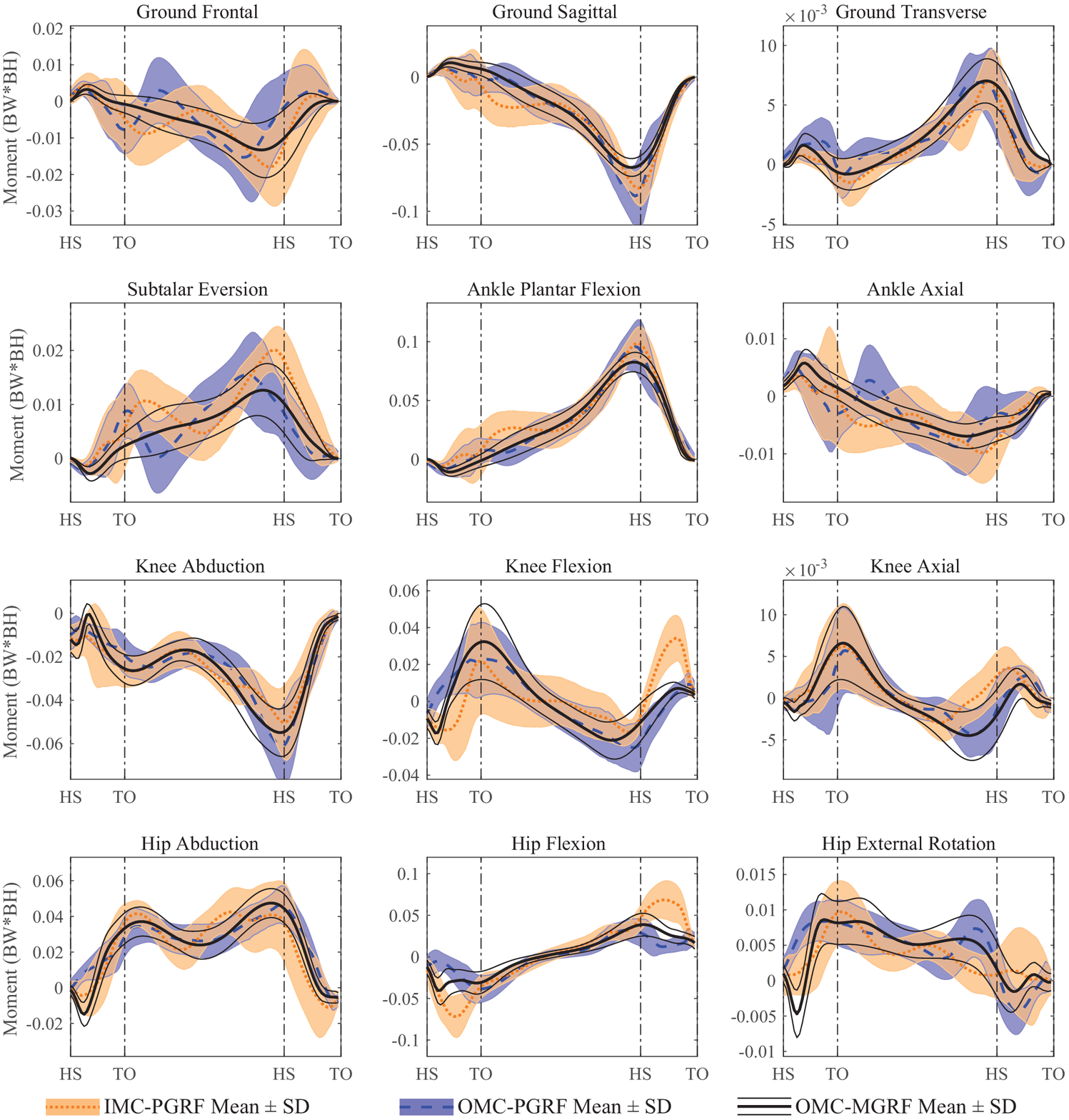}
\caption{Ground reaction and lower limb net joint moment estimates (standard deviation around mean) of the IMC-PGRF (orange shaded area around orange dotted line) and OMC-PGRF models (blue shaded area around blue dashed line) versus OMC-MGRF model (thin black solid lines around thick black solid line).}
\label{fig:Moment}
\end{figure}

The results of the accuracy analysis for GRF\&M and JRF\&M are presented in Table \ref{tab:IMC} and \ref{tab:OMC}, for IMC-PGRF and OMC-PGRF, respectively. The mean values and standard deviations of the curves from IMC-PGRF, OMC-PGRF, and OMC-MGRF models, are illustrated in Figures \ref{fig:Force} and \ref{fig:Moment}, for the forces and moments, respectively. 
\par
The Pearson correlation coefficients of the IMC-PGRF model were excellent for vertical ($\rho = 0.97$) and anteroposterior GRF\&M ($\rho = 0.91$) and strong for mediolateral GRF\&M ($\rho = 0.80$). For the same components, RMSD values observed were of $9.3 \pm 3.0$, $5.5 \pm 1.2$ and $2.1 \pm 0.6$ \%BW, respectively (mean $\pm$ standard deviation). The OMC-PGRF model performed better in the anteroposterior GRF\&M components ($\rho = 0.96$, RMSD $= 3.7 \pm 1.1$ \%BW), and similarly to IMC-PGRF for the other two GRF\&M components (mediolateral: $\rho = 0.79$, RMSD $= 1.9 \pm 0.5$ BW, vertical: $\rho = 0.99$, RMSD $= 5.9 \pm 1.9$ BW). 
\par
Concerning GRM, the sagittal plane was predicted with similar excellent correlations in both IMC-PGRF ($\rho = 0.91$) and OMC-PGRF ($\rho = 0.94$) driven models. The correlation coefficients for frontal and transverse GRM components found in the IMC-PGRF model were $\rho = 0.64$, $\rho = 0,82$, respectively, whereas in the OMC-PGRF model ($\rho = 0.66$, $\rho = 0,81$, respectively). The RMSDs found in the IMC-PGRF approach were $0.9 \pm 0.6 $, $ 1.6 \pm 0.6 $ ,  and $ 0.2 \pm 0.001$ \%BW*BH for frontal, sagittal and transverse GR\&M, respectively, which were either slightly higher or similar to the RMSDs of the OMC-PGRF approach ($0.7 \pm 0.2 $, $ 1.2 \pm 0.4 $,  and $ 0.2 \pm 0.1$ \%BW*BH, respectively). 

\begin{table}[H]
\centering
\footnotesize
\caption{IMC-PGRF-based ground and joint reaction forces (first three quantities) and net moments (second three quantities) versus OMC-MGRF. Pearson correlation coefficient is denoted with $\rho$. Absolute per body weight (or body weight times height) and relative root-mean-squared-difference are denoted with $RMSD$ (\%BW or \%BW*BH) and $rRMSD$ ($\%$), respectively. $M$ and $P$ indicate the magnitude and phase differences ($\%$).}
\label{tab:IMC}
\begin{tabular}{lrrrrr} 
\toprule
\multicolumn{1}{c}{} & \multicolumn{1}{c}{$\rho$} & \multicolumn{1}{c}{RMSD} & \multicolumn{1}{c}{rRMSD} & \multicolumn{1}{c}{M} & \multicolumn{1}{c}{P} \\ \midrule

\multicolumn{6}{l}{Ground}                                                             \\ \midrule
Anteroposterior   & 0.91        & 5.5 (1.2)    & 15.0 (2.4)  & -25.4 ( 7.3)  & 14.4 (3.2)    \\
Mediolateral      & 0.80        & 2.1 (0.6)    & 18.5 (3.2)  & 7.3 (19.3)    & 15.4 (3.8)    \\
Vertical          & 0.97        & 9.3 (3.0)    & 7.7 (2.1)   & -1.5 ( 1.5)   & 3.4 (1.0)     \\
Frontal           & 0.64        & 0.9 (0.6)    & 38.0 (23.1) & 125.5 (319.9) & 30.6 (17.3)   \\
Sagittal          & 0.91        & 1.6 (0.6)    & 17.5 ( 6.8) & 14.3 ( 18.2)  & 12.1 ( 4.5)   \\
Transverse        & 0.82        & 0.2 (0.1)    & 23.3 ( 7.2) & -8.5 ( 41.9)  & 17.8 ( 5.3)   \\ \midrule
\multicolumn{6}{l}{Ankle}                                                             \\ \midrule
Anteroposterior   & 0.84        & 22.2 (10.3)  & 26.1 (10.2) & 49.0 (45.8)   & 10.8 (2.1)    \\
Mediolateral      & 0.93        & 24.3 ( 8.9)  & 15.2 ( 5.3) & 14.3 (17.1)   & 7.9 (2.7)     \\
Proximodistal     & 0.93        & 88.5 (30.6)  & 13.6 ( 4.6) & 9.8 (14.1)    & 7.2 (2.3)     \\
Eversion          & 0.76        & 0.6 (0.2)    & 33.3 (20.2) & 107.7 (220.3) & 18.9 (10.7)   \\
Plantar Flexion   & 0.93        & 1.6 (0.6)    & 15.1 ( 6.6) & 10.6 ( 18.1)  & 9.9 ( 3.6)    \\
Axial             & 0.67        & 0.5 (0.2)    & 30.4 (12.2) & 46.5 ( 49.1)  & 27.2 (13.5)   \\ \midrule
\multicolumn{6}{l}{Knee}                                                             \\ \midrule
Anteroposterior   & 0.82        & 30.6 (10.3)  & 25.8 (9.7)  & 43.7 (53.5)   & 13.0 (4.5)    \\
Mediolateral      & 0.91        & 12.0 ( 3.5)  & 14.1 (3.8)  & 6.6 ( 8.6)    & 7.0 (2.0)     \\
Proximodistal     & 0.90         & 63.1 (26.9)  & 14.3 (6.6)  & 5.1 ( 9.1)    & 7.2 (2.8)     \\
Abduction           & 0.81        & 1.1 (0.4)    & 18.9 ( 6.8) & -2.7 (16.1)   & 10.7 ( 3.8)   \\
Flexion           & 0.58        & 1.9 (0.5)    & 29.8 ( 7.6) & 17.9 (45.0)   & 32.8 ( 9.6)   \\
Axial             & 0.73        & 0.3 (0.1)    & 25.4 (10.3) & 2.3 (30.5)    & 27.9 (13.8)   \\ \midrule
\multicolumn{6}{l}{Hip}                                                             \\ \midrule
Anteroposterior   & 0.71        & 17.6 ( 7.6)  & 27.2 (9.6)  & 6.8 (24.4)    & 27.6 (10.9)   \\
Mediolateral      & 0.73        & 27.0 (12.5)  & 23.0 (7.4)  & 7.7 (14.6)    & 10.6 ( 4.1)   \\
Proximodistal     & 0.78        & 102.8 (30.6) & 21.7 (4.5)  & 20.2 (10.0)   & 9.0 ( 2.5)    \\
Abduction         & 0.83        & 1.4 (0.7)    & 19.7 (5.8)  & 6.3 (16.9)    & 13.7 ( 7.9)   \\
Flexion           & 0.92        & 2.2 (0.6)    & 19.4 (4.2)  & 73.2 (26.3)   & 14.8 ( 4.2)   \\
External Rotation & 0.50         & 0.5 (0.2)    & 31.6 (6.6)  & -3.9 (36.4)   & 25.6 (10.1) \\
\bottomrule
\end{tabular}

\end{table}

\begin{table}[H]
\centering
\footnotesize
\caption{OMC-PGRF-based ground and joint reaction forces (first three quantities) and net moments (second three quantities) versus OMC-MGRF. Pearson correlation coefficient is denoted with $\rho$. Absolute per body weight (or body weight times height) and relative root-mean-squared-difference are denoted with $RMSD$ (\%BW or \%BW*BH) and $rRMSD$ ($\%$), respectively. $M$ and $P$ indicate the magnitude and phase differences ($\%$).}
\label{tab:OMC}
\begin{tabular}{lrrrrr}
\toprule
\multicolumn{1}{c}{} & \multicolumn{1}{c}{$\rho$} & \multicolumn{1}{c}{RMSD} & \multicolumn{1}{c}{rRMSD} & \multicolumn{1}{c}{M} & \multicolumn{1}{c}{P} \\ \midrule

\multicolumn{6}{l}{Ground}                                                            \\ \midrule
Anteroposterior   & 0.96        & 3.7 (1.1)   & 8.3 (2.0)  & 7.7 (12.0)   & 8.8 (1.8)     \\
Mediolateral      & 0.79        & 1.9 (0.5)   & 18.6 (4.1) & 2.4 (10.8)   & 15.2 (4.9)    \\
Vertical          & 0.99        & 5.9 (1.9)   & 4.9 (1.4)  & -1.2 ( 1.1)  & 2.1 (0.7)     \\
Frontal           & 0.66        & 0.7 (0.2)   & 30.3 (9.3) & 71.0 (122.2) & 24.5 (9.1)    \\
Sagittal          & 0.94        & 1.2 (0.4)   & 13.1 (3.8) & 15.9 ( 15.3) & 9.2 (3.2)     \\
Transverse        & 0.81        & 0.2 (0.1)   & 20.7 (7.5) & 7.1 ( 22.9)  & 17.5 (7.5)    \\ \midrule
\multicolumn{6}{l}{Ankle}                                                            \\ \midrule
Anteroposterior   & 0.83        & 18.9 ( 6.9) & 23.0 (6.1) & 37.3 (28.6)  & 10.8 (2.3)    \\
Mediolateral      & 0.96        & 16.1 ( 4.2) & 10.7 (2.6) & 6.8 ( 9.6)   & 5.8 (2.1)     \\
Proximodistal     & 0.96        & 62.2 (17.6) & 9.8 (2.7)  & 7.1 ( 9.0)   & 5.2 (1.8)     \\
Eversion          & 0.76        & 0.5 (0.1)   & 25.5 (7.0) & 45.3 (64.1)  & 18.7 (10.2)   \\
Plantar Flexion   & 0.96        & 1.0 (0.3)   & 10.1 (3.3) & 5.9 (10.0)   & 7.0 ( 2.6)    \\
Axial             & 0.64        & 0.5 (0.1)   & 27.2 (7.3) & 33.3 (36.9)  & 27.5 (11.5)   \\ \midrule
\multicolumn{6}{l}{Knee}                                                            \\ \midrule
Anteroposterior   & 0.93        & 11.9 ( 4.5) & 12.3 (4.3) & -7.3 (8.7)   & 7.4 (2.0)     \\
Mediolateral      & 0.96        & 7.2 ( 2.0)  & 8.8 (2.6)  & -4.2 (5.6)   & 4.4 (1.0)     \\
Proximodistal     & 0.95        & 41.7 (12.0) & 9.3 (2.6)  & -2.7 (5.8)   & 4.9 (1.2)     \\
Abduction           & 0.91        & 0.8 (0.2)   & 12.6 (2.6) & -0.1 (10.5)  & 7.7 (1.6)     \\
Flexion           & 0.86        & 0.9 (0.3)   & 16.7 (4.8) & -1.7 (14.3)  & 16.9 (5.2)    \\
Axial             & 0.82        & 0.2 (0.1)   & 18.5 (6.6) & -3.4 (17.7)  & 20.6 (8.0)    \\ \midrule
\multicolumn{6}{l}{Hip}                                                            \\ \midrule
Anteroposterior   & 0.89        & 9.9 ( 3.6)  & 16.0 (6.7) & -10.4 (10.6) & 16.6 (7.6)    \\
Mediolateral      & 0.92        & 14.7 ( 4.0) & 12.7 (3.1) & -1.9 ( 6.9)  & 6.2 (1.5)     \\
Proximodistal     & 0.92        & 50.0 (15.9) & 11.5 (2.6) & -4.6 ( 6.1)  & 5.5 (1.2)     \\
Abduction         & 0.91        & 0.8 (0.2)   & 13.3 (2.6) & -3.2 ( 6.3)  & 8.7 (2.4)     \\
Flexion           & 0.86        & 1.3 (0.4)   & 16.4 (3.4) & -9.3 (12.3)  & 18.0 (4.1)    \\
External Rotation & 0.68        & 0.3 (0.1)   & 22.5 (3.7) & 6.5 (15.8)   & 18.8 (4.8)   \\
\bottomrule
\end{tabular}

\end{table}

\section{Discussion}
We have presented a method to perform musculoskeletal-based inverse dynamics using exclusively IMC input (IMC-PGRF). First, we compared the kinematic joint angle estimates of the lower limbs against those assessed through a conventional, laboratory-based OMC input. In addition, we tested the performance of the approach in calculating the JRF\&M, while predicting the GRF\&M from the kinematics, against a similarly constructed model (OMC-MGRF) which uses input from both FP and OMC. Finally, we performed a similar comparison to evaluate the predicted kinetics of a model driven exclusively by OMC input (OMC-PGRF).
\par
Regarding the IMC-based joint angles in the musculoskeletal model, all three sagittal plane angles provided excellent correlations (range: 0.95-0.99) and average RMSD values remained below $6^{\circ}$. Slightly lower correlations were observed in the frontal and transverse plane angles, which can be explained due to the smaller range of motion within these planes. For instance, even though the hip abduction and external rotation joint angles present absolute RMSD values similar to the flexion component, their rRMSDs which take into account the range of motion are two and three times higher, respectively. 
\par
Both GRF\&M and JRF\&M of the vertical axis presented higher correlations and lower RMSDs than the ones in the anteroposterior and mediolateral axes. Similarly, sagittal plane moments were found in most cases to be more accurate than frontal and transverse plane moments. By visual inspection of the curves, we observe that the magnitude of the IMC-PGRF anteroposterior GRF\&M seems to be underestimated both in the negative early stance and positive late stance peak, which can be confirmed by the magnitude difference for that curve ($M=-28.3\%$). However, this behaviour is not observed in the OMC-PGRF, nor during the single stance of the IMC-PGRF curve. Despite the higher rRMSD found in the non-sagittal joint angles, the performance of the IMC-PGRF in the mediolateral, frontal and transverse plane GRF\&M components matched closely the OMC-PGRF approach. This observation reveals that OMC-based kinematics suffer from errors of similar size, when capturing the typically small movements of the frontal and transverse planes, given the fact that both IMC-PGRF and OMC-PGRF had the same model characteristics. Therefore, OMC-MGRF should also be used with caution, when comparing either kinematic or JRF\&M quantities of the non-sagittal planes.  
\par
A number of error sources contribute to discrepancies in the OMC kinematics. First, soft tissue artefacts can create a relative movement of the marker with respect to the bone  \citep{Chiari2005226,Leardini2005212}. In addition, mismatches between the experimental and modelled marker positions can lead to errors in segment orientations calculated during inverse kinematics. Both error sources would have a relatively larger impact on the kinematics of the frontal and transverse plane, than on the sagittal plane. Finally, the JRF\&M of the OMC-PGRF were compared against a non-independent OMC-MGRF reference, which could have contributed to underestimation of the actual errors.
\par
The IMC-PGRF approach has a number of possible sources of errors which would influence the performance. Errors in segment kinematics may stem due to the N-pose calibration assumptions. In particular, mismatches between the practised and modelled N-pose could result in offsets in the estimated positions. Other common error sources in IMC include manual measurements of segment lengths as well as IMU inaccuracies. In addition, the stick figure model, which was utilized to recreate the VMs, has a higher number of DOF, compared to the musculoskeletal model used. 
\par
A possible source of error in all inverse dynamic approaches concerns the inertial parameters (masses and moments of inertia), as well as the center of mass (CoM) locations of each human body segment, which were calculated based on anthropometric tables found in the literature. 
\par
Future work could evaluate the muscle, bone, or ligament force estimates \citep{Marra2015}. Moreover, reducing the number of IMUs to assess kinetic quantities could be investigated \citep{Giuberti2014,Wouda2016}. Since this study was limited to gait-related movements, it would be of interest to validate it during activities of daily living \citep{Fluit20142321} or sports activities \citep{Skals2016}, as well as other groups, including patients with musculoskeletal disorders, elderly, or obese subjects.  
\par
\section{Conclusion}
In this study, we have demonstrated that the prediction of GRF\&M as well as JRF\&M using musculoskeletal-model-based inverse dynamics based on only IMC data provides comparable performance to both OMC-PGRF and OMC-MGRF methods. The proposed method allows assessment of kinetic variables outside the laboratory.  

\section{Conflicts of interest}
Three of the authors are employees of Xsens Technologies B.V. that manufactures and sells the Xsens MVN. One of the authors is employee of AnyBody Technology A/S that owns and sells the AnyBody Modeling System.  
\section{Acknowledgements}
This study was performed in the context of KNEEMO Initial Training Network, funded by the European Union’s Seventh Framework Programme for research, technological development, and demonstration under Grant Agreement No. 607510 (www.kneemo.eu). This work was also supported by the Danish Council for Independent Research under grant no. DFF-4184-00018 to M. S. Andersen. Finally, this research received funding in part from the European Union’s Horizon 2020 research and innovation programme under grant agreement No. 680754 (The MovAiD project, www.movaid.eu).

\pagebreak
\section*{References}

\section*{Copyright Information}

\begin{figure}[H]	
	\begin{minipage}[c]{0.2\textwidth}
		\href{http://creativecommons.org/licenses/by-nc-nd/4.0/}{\includegraphics[scale=0.7]{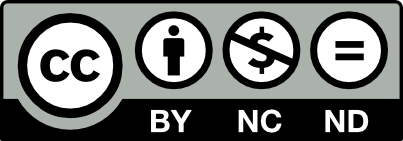}}		
	\end{minipage}\hfill
	\begin{minipage}[c]{0.78\textwidth}
		\caption*{\href{http://creativecommons.org/licenses/by-nc-nd/4.0/}{{\textcopyright} 2017. This manuscript version is made available under the CC-BY-NC-ND 4.0 license http://creativecommons.org/licenses/by-nc-nd/4.0/}
		} \label{cc-by-nc-nd}
	\end{minipage}
\end{figure}

\end{document}